\documentclass[%
 reprint,
nofootinbib,
 amsmath,amssymb,
 aps,
]{revtex4-2}

\usepackage{graphicx}
\usepackage{dcolumn}
\usepackage{bm}
\usepackage{xcolor}
\usepackage{physics}
\usepackage{amsthm}
\usepackage{comment}
\usepackage{float}
\usepackage{dsfont}
\usepackage[colorlinks=true,bookmarks=false,citecolor=blue,urlcolor=blue,linkcolor=blue]{hyperref}

\begin{document}

\renewcommand{\section}[1]{\emph{#1}.---}
\newcommand{\ie}{\emph{i.e.}}
\newcommand{\eg}{\emph{e.g.}}
\newcommand{\cf}{\emph{cf.}}
\newcommand{\ea}{\emph{et al.}}
\newcommand{\etc}{\emph{etc.}}

\preprint{APS/123-QED}

\title{No Tradeoff between Coherence and Sub-Poissonianity for Heisenberg-Limited Lasers}

\author{L.~A.~Ostrowski}
 \email{lucas.ostrowski@griffithuni.edu.au}
\author{T.~J.~Baker}
\author{S.~N.~Saadatmand}
\author{H.~M.~Wiseman}
 \email{h.wiseman@griffith.edu.au}
\affiliation{%
 Centre for Quantum Dynamics, Griffith University, Yuggera Country, Brisbane, Queensland 4111, Australia
}%

\date{\today}

\begin{abstract}
The Heisenberg limit to laser coherence $\mathfrak{C}$—the number of photons in the maximally populated mode of the laser beam—is the fourth power of the number of excitations inside the laser. We generalize the previous proof of this upper bound scaling by dropping the requirement that the beam photon statistics be Poissonian (i.e., Mandel’s $Q=0$).  We then show that the relation between $\mathfrak{C}$ and sub-Poissonianity ($Q<0$) is win-win, not a tradeoff. For both regular (non-Markovian) pumping with semi-unitary gain (which allows $Q\xrightarrow{}-1$), and random (Markovian) pumping with optimized gain, $\mathfrak{C}$ is maximized when $Q$ is minimized.
\end{abstract}

\maketitle

\section{Introduction} The characteristic ability of a laser to produce optical fields with a high degree of coherence has led to their widespread application in precision technology. This is especially true in the quantum regime, as the ability to access such fields has directly permitted quantum optics to flourish~\cite{Wineland, Bloch, Deleglise, Hensen, Minev}, paving the way for the advent of quantum technology~\cite{Cirac, qoptics_to_qinfo}. This defining feature of coherence can be quantified for a continuous-wave laser by $\mathfrak{C}$, the mean number of photons in a maximally populated mode of the beam~\cite{HL}. For a laser, unlike other light sources, this number will typically be enormous, and may be far greater than the number of excitations stored inside the laser itself. Considering a laser beam with ``ideal" properties---being describable by a coherent state undergoing pure phase-diffusion~\cite{SSLamb, Louisell, Carmichael99}---$\mathfrak{C}$ can be intuitively expressed in terms of the photon flux from the laser, $\mathcal{N}$, and its linewidth, $\ell$, as $\mathfrak{C} = 4\mathcal{N}/\ell$. 

The foundations of laser theory was the derivation, by Schawlow and Townes, of the laser linewidth, $\ell$~\cite{SQL}. This formula sets a bound on $\ell$, and hence $\mathfrak{C}$ in terms of an energy resource, namely, $\mathfrak{C}_{\textrm{STL}} = \Theta(\mu^2)$~\cite{Wiseman1999}.
Here, $\mu$ is the mean number of \textit{total} coherent excitations within the laser (including photons, atomic excitations, excitons, etc). This limit was long thought to have been an absolute limit, but recent works have shown that this may be greatly surpassed~\cite{HL, Pekker}. In particular, the authors of Ref.~\cite{HL} considered the Heisenberg limit (HL) for $\mathfrak{C}$, that is, the ultimate achievable limit that is imposed by quantum mechanics, rather than by standard technologies. They proved rigorously, under some natural conditions, that the HL is
\begin{align}\label{coh_HL}
    \mathfrak{C}_{\textrm{HL}} = \Theta(\mu^4).
\end{align}
These conditions conceive of a laser as a device that produces an output closely approximating an ideal laser beam without restricting the processes by which it achieves this, other than that the phase of the beam proceeds \textit{only} from the phase of the excitations stored within the device. The key to attaining this quadratic quantum enhancement is to change both the pumping and output coupling of the laser to be highly nonlinear, a feat that is potentially achievable in circuit-QED~\cite{HL, Pekker}.

This letter advances the physics objectives within this field in two ways. First, we generalize the salient result of Ref.~\cite{HL}, such that the HL scaling $\mathfrak{C}=O(\mu^4)$ is re-derived under significantly relaxed conditions on the beam. This is motivated by the prospects of experimental realization. The strict conditions placed on the beam given in Ref.~\cite{HL} will probably not be satisfied in the most feasible near-term experimental hardware that would surpass the standard quantum limit for $\mathfrak{C}$~\cite{Ostrowski}.

Second, this relaxation permits us study the photon statistics of Heisenberg-limited lasers (\ie, those achieving Eq.~(\ref{coh_HL})). This has furnished us with a fundamental insight to the nature of laser radiation. That is because the more general HL derived here applies to laser beams exhibiting \textit{sub-Poissonian} photon statistics, even allowing the number fluctuations to vanish for long counting intervals. Interest in the production of sub-Poissonian light from lasers started in the 1980s and early 1990s~\cite{Golubev1984,Yamamoto1986a,Machida1987,Bergou1989,Richardson1990,Walls1990,Wiseman1991,Ralph1991a,Ralph1991b,Yamamoto1992}. It has remained an active area of research since then~\cite{Choi2006,Koppenhofer2016,Koppenhofer2017,Canela}, due to the foreseeable broad application in areas such as quantum-enhanced measurement, communication, sensing and information processing~\cite{Kolobov,Yamamoto1986b,Polzik,Ralph1995,Davidovitch,Golubeva2008,Walls_Milb,Korolev2019,Mork2020,Goldberg,Hosseinidehaj2022,Zhao2022}.

Among the early work within this area, several theoretical studies were conducted to determine if a tradeoff existed between the coherence and degree of sub-Poissonianity in \textit{standard} laser beams (\ie, those which achieve $\mathfrak{C}=\Theta(\mu^2)$ at best). It was shown that for a laser with standard output coupling, pumping which achieves a sub-Poissonian output does not necessarily significantly increase the phase diffusion rate~\cite{Bergou,Benkert_Phase,wiseman1993,Drummond} (though in other models it does~\cite{Wiseman1991}). In this work, we extend the results from these early studies to the extreme case, to laser models which exhibit a phase-diffusion rate that is as small as permitted by the laws of quantum mechanics. 

We show that there is no tradeoff between coherence and the degree of sub-Poissonianity (quantified by the Mandel-$Q$ parameter~\cite{Mandel}) in Heisenberg-limited laser beams. Moreover, a ``win-win situation" occurs in both of the families of laser models we introduce. These models exhibit Heisenberg-limited coherence as well as sub-Poissonian beam photon statistics in certain parameter regimes, and there is a perfect correlation between an increase in the coherence and a decrease in the Mandel-$Q$ parameter of the beam. An important point to stress here is that ``external squeezing" is not an alternative to our models. The conception of a ``laser" in Refs.~\cite{HL,Pekker} and this work means considering the whole device, encompassing all processes that give rise to both the coherence and the sub-Poissonianity in the beam, as a laser. Throughout, we refer to the Companion Paper~\cite{Ostrowski} for details of proofs and much more.

\section{Beam Coherence and Sub-Poissonianity} 
We start with the two quantities relating to a laser beam that are the focus of this Letter. The first of these is \textit{laser coherence}, $\mathfrak{C}$, which relates to the phase fluctuations of the beam. A unidirectional beam of light produced by a laser can be aptly described as a scalar quantum field with the single-parameter field annihilation operator, $\hat{b}(t)$. For such a field, $\mathfrak{C}$ is defined as the maximally populated mode (within some frequency band, if required)~\cite{HL},
\begin{align}\label{coh_general}
    \mathfrak{C} := \max_{u\in\mathfrak{u}}\langle \hat{b}_u^\dagger \hat{b}_u \rangle.
\end{align}
Here, $\hat{b}_u = (1/\sqrt{I_u})\int_{-\infty}^\infty dt u(t)\hat{b}(t)$ defines the annihilation operator for mode $u$, which is normalized such that $I_u = \int_{-\infty}^\infty dt|u(t)|^2$.

We can gain intuition about this quantity by analyzing Eq.~(\ref{coh_general}) under some additional assumptions that are characteristic of an `ideal' laser beam. First, for a beam with translationally invariant statistics, $\mathfrak{C}/2\pi$ is simply the peak of the power spectrum $P(\omega)$, the Fourier transform of the correlation function $\langle\hat{b}^\dagger(t+\tau)\hat{b}(t)\rangle$~\cite{HL}. Additionally, the early work on laser theory during the 1960's and 70's~\cite{SSLamb, Louisell} showed that, when technical noise is negligible, the state of a laser beam can be well-approximated by the eigenstate
\begin{align}\label{las_ideal}
    \ket{\beta(t)} = \ket{\sqrt{\mathcal N}e^{i\sqrt{\ell}W(t)}},
\end{align}
of $\hat{b}(t)$ for each $t$, where $W(t)$ represents a Wiener process. That is, it is a coherent state undergoing pure phase-diffusion at rate $\ell$. For such a state, the photon statistics are Poissonian~\cite{SSLamb,Arecchi} and the power spectrum is Lorentzian,
\begin{align}\label{P_spec}
    P(\omega) = \frac{\mathfrak{C}}{2\pi}\frac{(\ell/2)^2}{(\omega-\omega_0)^2+(\ell/2)^2},
\end{align}
where $\omega_0$ is the central frequency. Since the linewidth of this ideal beam is entirely due to phase-diffusion, the coherence time, $1/\ell$, can be thought of as the time for the phase of the laser to become fairly randomized. Multiplying this by the photon flux, $\mathcal{N}=\int d\omega P(\omega)$, one has $\mathcal{N}/\ell = \mathfrak{C}/4$. In the context of such an ideal laser, $\mathfrak{C}$ may therefore be interpreted roughly as \textit{the number of photons emitted into the beam with a well-defined phase relationship}.

Second, to characterize the intensity fluctuations in the beam, we employ the Mandel-$Q$ parameter~\cite{Mandel} defined on the output field over the duration $T$:
\begin{align}\label{mandel}
    Q_T := \frac{\langle (\Delta \hat{n}_T)^2 \rangle - \langle\hat{n}_T\rangle}{\langle\hat{n}_T\rangle},
\end{align}
with $\hat{n}_T \equiv \int_{T_0}^{T_0+T}ds\hat{b}^\dagger(s)\hat{b}(s)$ as the number operator for the section of the beam over the interval $(T_0,T_0+T]$. Note that $Q_T$ does not depend on $T_0$ for a stationary field. This is an affine function of the \textit{variance in the number of photon detections made by an ideal detector monitoring the beam}. It is defined so that $Q\in[-1,0)$ implies sub-Poissonian photon statistics (a variance less than the mean) in the beam over the interval $(T_0,T_0+T]$. Here we consider the limit of long counting intervals, $T\xrightarrow{}\infty$. That is, the measure of beam intensity fluctuations we seek to minimize is $Q\equiv Q_{T\xrightarrow{}\infty}$.

\section{Generalizing the Heisenberg Limit for $\mathfrak{C}$}
In Ref.~\cite{HL} four conditions were considered for a laser and the beam it produces. For a device satisfying these conditions, it was shown that $\mathfrak{C}\lesssim2.9748\mu^4$, representing a Heisenberg limit for the coherence of its beam. One of these conditions implies a beam that exhibits $Q_T\approx0$. Here we summarize these four conditions and the motivation behind them, before showing how they can be modified such that a more general Heisenberg limit for $\mathfrak{C}$ can be derived. This generalized limit applies to lasers that produce a beam for which its photon statistics can have a significant degree of sub-Poissonianity. The original four conditions are as follows:

Condition 1, or \textit{One Dimensional Beam}, states that the beam propagates away from the laser in one direction at a constant speed, occupying a single transverse mode and polarization. This allows the output of the device to be described by a scalar quantum field with the annihilation operator $\hat{b}(t)$ satisfying $[\hat{b}(s),\hat{b}^\dagger(t)]=\delta(s-t)$. 

Condition 2, or \textit{Endogenous Phase}, requires that coherence in the beam proceeds only from within the laser. This is satisfied if a phase shift imposed on the laser state at some time $T_0$ will lead, in the future, to the same phase shift on the beam emitted after time $T_0$, as well as on the laser state. When considering specific laser models in practice, this follows naturally if the mechanisms of gain and loss of excitations in the laser correspond to $U(1)$-covariant processes~\cite{HL,Bartlett}. 

Condition 3, or \textit{Stationarity}, requires the laser and the beam it produces to have a long time limit that is unique and invariant under time translation. This means that the mean excitation number within the laser, $\langle \hat{n}_{\rm c} \rangle$, has a unique stationary value $\mu$. 

Condition 4, or \textit{Ideal Glauber$^{(1),(2)}$-Coherence}, states that the stationary first- and second-order Glauber coherence functions~\cite{Glauber} of the laser beam---$g_{\rm laser}^{(1)}(s,t)$ and $g_{\rm laser}^{(2)}(s,s',t',t)$, respectively---\textit{well-approximate} those of an ideal standard laser beam state given by Eq.~(\ref{las_ideal}). The motivation for this condition is that it allows the beam to exhibit the typical properties of an ideal laser beam to a good approximation (such as a Lorentzian power spectrum and Poissonian beam statistics), while also allowing the upper bound on $\mathfrak{C}$ to be obtained by the specific method of proof used in Ref~\cite{HL}.

In this work, we adopt Conditions 1--3 outright as requirements for a laser to satisfy, however Condition 4 is significantly relaxed to encompass a much more general range of laser models. The new Condition 4 that we consider, \textit{Passably Ideal Glauber$^{(1),(2)}$ Coherence}, requires bounding the differences between the Glauber coherence functions (first- and second-order) for the laser beam in question and an ideal beam as follows:
\begin{subequations}\label{c4}
    \begin{align}\label{c4.1}
        |g_{\rm laser}^{(1)}(s,t)-g^{(1)}_{\rm ideal}(s,t)|= O(1),
    \end{align}
    \begin{align}\label{c4.2}
        |g_{\rm laser}^{(2)}(s,s',t',t)-g^{(2)}_{\rm ideal}(s,s',t',t)|= O(\mathfrak{C}^{-1/2}),
    \end{align}
\end{subequations}
for all values of the time arguments such that the difference between any two times is $O(\sqrt{\mathfrak{C}}/\mathcal{N})$. 

The original Condition 4 of Ref.~\cite{HL} is the same as the one above, except for having $o$ in place of $O$ in Eqs.~(\ref{c4}a)~and~(\ref{c4}b). As we will show, this revised condition now allows for models that produce highly sub-Poissonian beams, where $Q$ approaches its minimum of $-1$, corresponding to vanishing photon noise in the beam for long counting intervals. Specifically, this is permitted because of the weaker constraint on $g^{(2)}_{\rm laser}(s,s',t',t)$. 

We note here that because $g^{(1)}_{\rm laser}(s,t)$ is relatively unconstrained by Eq.~(\ref{c4.1}), the interpretation provided for $\mathfrak{C}$ below Eq.~(\ref{P_spec}) would not generally be vaild for a laser device satisfying this criteria. However, for the laser models presented in this work, $g^{(1)}(s,t)$ \textit{does} exhibit an exponential decay characteristic of an ideal beam to a very good approximation~\cite{Ostrowski}. Thus we can still think of $\mathfrak{C}$ as the number of photons emitted in a coherence time.

A proof for the upper bound on $\mathfrak{C}$ for a beam satisfying these more relaxed conditions is now outlined; for details, see Ref.~\cite{Ostrowski}.

\textbf{Theorem 1:} (Generalisation of the upper bound on $\mathfrak{C}$ for sub-Poissonian lasers). \textit{For a laser which satisfies Conditions 1--3, along with the new Condition 4, stated above, the coherence is bounded from above:}
\begin{align}\label{Heisenberg}
    \mathfrak{C} = O(\mu^4),
\end{align}
\textit{with $\mu$, the mean number of excitations within the laser.}

The methodology applied to show this follows closely that of Ref.~\cite{HL}, which boils down to comparing the precision between different methods of estimating the optical phase of a laser. In particular, we consider a laser model satisfying Conditions 1--3, and the Passably Ideal Glauber Coherence Condition, operating at steady state. One observer, Effie, then encodes a random phase, $\phi_F$, onto the laser state by performing a \textit{filtering} heterodyne measurement on its beam over the time interval $[T-\tau,T)$. 

It is the job of a second observer, Rod, to estimate this encoded optical phase. we consider two methods to achieve this. The first is a \textit{retrofiltering} heterodyne measurement of the beam over the interval $(T,T+\tau]$. From the theory of heterodyne detection~\cite{WisemanMilburn}, it is possible to quantify how correlated Rod's retrofiltering measurement outcome, $\phi_R$, is with Effie's result, $\phi_F$. Considering the relative difference, in the deviations of the estimates, between the laser model and that of an ideal beam~\cite{Ostrowski}, we arrive at the expression
\begin{align}\label{hetero ub}
    1 - |\langle e^{i(\phi_R - \phi_F)}\rangle|^2=O(\mathfrak{C}^{-1/2}),
\end{align}
for the choice $\tau = \Theta(\sqrt{\mathfrak{C}}/\mathcal{N})$, which minimizes the scaling of the mean-square error (MSE) in Rod's measurement~\cite{HL}. It should be recognized that the LHS of this equation provides a measure of phase spread, as for small errors $\theta$, $1-|\langle e^{i\theta} \rangle|\approx\langle\theta^2\rangle - \langle\theta\rangle^2$ represents the MSE. 

Equation~(\ref{hetero ub}) highlights a key difference here with the result of Ref.~\cite{HL}; there, a specific prefactor was able to be obtained in Eq.~(\ref{hetero ub}). Because of our relaxed constraints imposed on the beam by our fourth Condition, a prefactor for this quantity, and hence for the upper bound on $\mathfrak{C}$, is no longer able to be attained. Regardless of this, we are still able to derive a limit for the scaling of $\mathfrak{C}$ with $\mu$, which is sufficient to talk of a Heisenberg limit.

To derive this, consider Rod's second method for obtaining an estimate for $\phi_F$: performing a direct phase measurement on the laser device at time $T$. It is a known result~\cite{Bandilla} that the MSE of such an estimate, $\phi_D$, is bounded below, asymptotically,
\begin{align}\label{bound3}
    1 - |\langle e^{i(\phi_D - \phi_F)} \rangle|^2 = \Omega(\mu^{-2}).
\end{align}
Finally, $\phi_R$ obtained via the retrofiltering measurement outlined above cannot outperform $\phi_D$ as an estimate of the encoded laser phase $\phi_F$~\cite{HL}. It therefore follows from Eqs.~(\ref{hetero ub})~and~(\ref{bound3}) that $\mathfrak{C} = O(\mu^4)$ represents the Heisenberg limit for any laser satisfying the stated conditions. From our adopted method of proof, these are the loosest requirements on the laser for which this $\mu^4$ scaling can be proven to be the HL.

\section{Sub-Poissonian, Heisenberg-Limited Laser Models}
\begin{figure}[H]
\includegraphics[width=1.0\columnwidth]{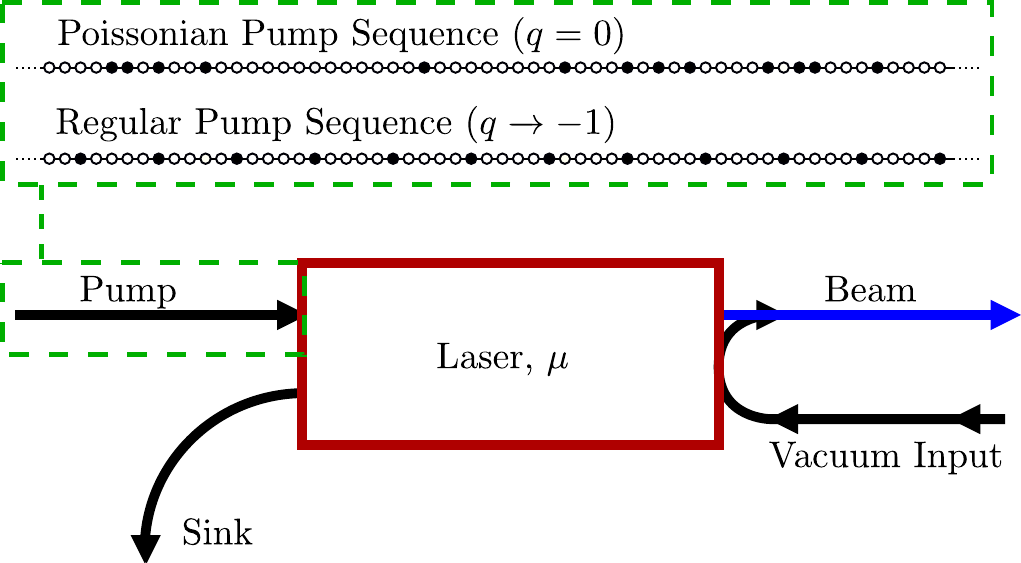}
\caption{\label{schematic1} Basic schematic of a laser showing the key components of the two families of models discussed in the text. Top green box depicts different statistics for the pump, applicable to the $q$-family, while the $\lambda$-family is constrained to have Poissonian pumping statistics.}
\end{figure}

We now present two families of laser models, both satisfying our four conditions~\cite{Ostrowski}, and both exhibiting Heisenberg-limited coherence with $\mathfrak{C}=\Theta(\mu^4)$. Fig.~\ref{schematic1} shows the key components of these: the laser ``cavity" (a $D$-level system with the non-degenerate number operator $\hat{n}_c=\sum_{n=0}^{D-1}n\ketbra{n}{n}$) storing an average of $\mu$ excitations at steady state, a pump that adds incoherent excitations into the cavity, a vacuum input, which, upon reflection, becomes the beam, and finally a sink for all excitations that leave the cavity but not in the beam.

The two families have been developed by making generalizations to the original laser model of Ref.~\cite{HL}, with particular regard to the pumping mechanisms. In a frame rotating at cavity resonance, the dynamics of the laser cavity for this original model can be written in terms of a master equation in Lindblad form $\dot{\rho} = \mathcal{D}[\hat{G}]\rho + \mathcal{D}[\hat{L}]\rho$, where $\mathcal{D}[\hat{c}]\bullet:=\hat{c}\bullet\hat{c}^\dagger-(1/2)\{\hat{c}^\dagger\hat{c},\bullet\}$ is the usual Lindblad superoperator~\cite{WisemanMilburn}. Here, $\hat{G}$ and $\hat{L}$ are the single-excitation gain and loss operators, which, in the number basis, have non-zero matrix elements $\hat{G}_{n,n-1}\propto 1$ and $\hat{L}_{n-1,n}\propto (\rho_{n-1}/\rho_n)^{1/2}$. Here $\rho_n=\bra{n}\rho_{\rm ss}\ket{n}$ are the non-zero elements of the cavity steady state,
\begin{align}\label{cav_dist}
    \rho_n \propto \sin^4\left({\pi\frac{n+1}{D+1}}\right), \quad 0\leq n\leq D-1.
\end{align}
In this scenario, we say that the gain operator is \textit{quasi-isometric} ($\hat{G}^\dagger\hat{G} = \mathds{1} - \ket{D-1}\bra{D-1}$). This laser model produces a beam that exhibits Poissonian beam photon statistics, in accordance with the strict Ideal Glauber$^{(1),(2)}$-Coherence condition of Ref.~\cite{HL}. The ansatz for the cavity steady state~(\ref{cav_dist}) was suggested by numerical optimization of $\mathfrak{C}$ based on infinite matrix product state (iMPS) techniques~\cite{HL,Saadatmand}.

The first family of models that we introduce can be described with a master equation in the same form as that described above, while making the substitutions $\hat{G}\xrightarrow{}\hat{G}^{(\lambda)}$ and $\hat{L}\xrightarrow{}\hat{L}^{(\lambda)}$. Here, the parameter $\lambda\in\mathbb{R}$ generalizes the gain and loss operators according to 
\begin{align}\label{gainloss}
    G^{(\lambda)}_{n,n-1} \propto \left(\rho_n/\rho_{n-1}\right)^{\lambda/2}, \quad L^{(\lambda)}_{n-1,n} \propto \left(\rho_{n-1}/\rho_n\right)^{(1-\lambda)/2}.
\end{align}
This preserves the same ansatz for the cavity steady state~(\ref{cav_dist}) as in Ref.~\cite{HL}, and the original model corresponds to $\lambda=0$ (flat gain~\cite{Wiseman1999}). This \textit{``$\lambda$-family"} of laser models generally describes a \textit{randomly-pumped} gain mechanism, which, for $\lambda\neq0$, is \textit{non-isometric} ($\hat{G}^\dagger\hat{G}$ is far from the identity $\mathds{1}$)~\cite{Ostrowski}.

The second family that we introduce instead describes a \textit{regularly pumped (non-Markovian)} gain mechanism, which is \textit{quasi-isometric}. The cavity dynamics for this \textit{``$q$-family"} can be approximated as
\begin{align}\label{mastereqq}
    \dot{\rho} = \left(\mathcal{D}[\hat{G}^{(0)}] +\frac{q}{2}\mathcal{D}[\hat{G}^{(0)}]^2 + \mathcal{D}[\hat{L}^{(-q/2)}]\right)\rho,
\end{align}
where $q\in(-1,\infty]$ represents the Mandel-$Q$ parameter of the pumping statistics (see Fig.~\ref{schematic1}). This also has the steady state~(\ref{cav_dist}) in the limit $D\xrightarrow{}\infty$~\cite{Ostrowski} and the loss operator in Eq.~(\ref{mastereqq}) is as defined in Eq.~(\ref{gainloss}). Setting $q=0$ thus reduces Eq.~(\ref{mastereqq}) to the model presented in Ref.~\cite{HL}. This master equation is only an approximation because it is a Markovian equation describing a generally non-Markovian process. Regardless of this, master equations of this form have long been employed to model regularly-pumped lasers~\cite{Golubev1984,haaketanwalls,Bergou,wiseman1993,Walls_Milb} and the results that it yields are physically reasonable. For details, see Ref.~\cite{Ostrowski}.

\begin{figure}[H]
\includegraphics[width=1.0\columnwidth]{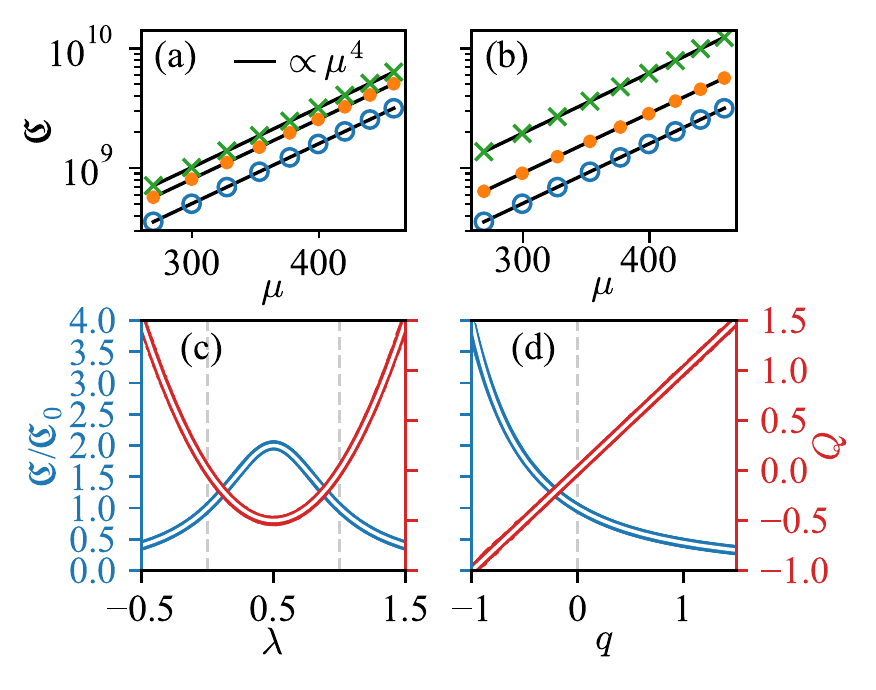}
\caption{\label{fig:2} (a): iMPS calculations of beam coherence $\mathfrak{C}$ for the $\lambda$-family of laser models as a function of average cavity excitation number $\mu$. Circles, dots, and crosses correspond to $\lambda = 0.0$, $0.25$, and $0.5$, respectively. Solid lines depict fitted power laws to the data, indicating a scaling of $\mathfrak{C}=\Theta(\mu^4)$. (b): Same as (a), but for the $q$-family of laser models, with $q = 0.0$, $-0.5$, and $-1.0$, respectively. (c,d): iMPS calculations of $\mathfrak{C}$ (normalized to $\mathfrak{C}_0=\mathfrak{C}_{\lambda=0}=\mathfrak{C}_{q=0}$) and $Q$-parameter for the $\lambda$- and $q$-families, respectively, for a fixed $\mu = 499.5$. White overlapped curves (often appearing centered over the numerics) are analytical formulae discussed in the text. Vertical grey lines highlight the model parameter values which give rise to Poissonian beam photon statistics.}
\end{figure}

Figures~\ref{fig:2}a--b show the scaling of $\mathfrak{C}$, for members of both the above families, with the mean photon number $\mu=(D-1)/2$. These quantities were numerically computed by discretizing the output beam and treating the laser as an iMPS sequential quantum factory~\cite{schon1,schon2}; see also Refs.~\cite{HL,Ostrowski}. Fitting a power law to each of these curves reveals $\mathfrak{C}=\Theta(\mu^4)$ for every choice of $\lambda$ and $q$, thus saturating the upper bound scaling of Eq.~(\ref{Heisenberg}). These plots also indicate that the prefactor of these power laws are optimized by the choice $\lambda=0.5$ in the $\lambda$-family, and $q=-1$ in the $q$-family. These prefactors are roughly, respectively, a factor of $2$ and $4$ larger than that of the model from Ref.~\cite{HL}.

This interesting detail is investigated in more depth in Figures~\ref{fig:2}d--e, plotting $\mathfrak{C}$ and $Q$ for both families of laser models as functions of $\lambda$ and $q$, for fixed $D=1000$. We find that increasing beam coherence is perfectly correlated with a reduction in beam photon noise. For the $\lambda$-family, $\mathfrak{C}_\lambda\approx\mathfrak{C}_{\lambda=0}/[2(\lambda-1/2)^2+1/2]$, while $Q_\lambda\approx2\lambda(\lambda-1)$. For the $q$-family, $\mathfrak{C}_q\approx\mathfrak{C}_{q=0}/(1+q/2)^2$, while $Q_q\approx q$, mirroring the Mandel-$Q$ parameter of the pumping mechanism. These formulae are indicated by the overlapped white curves and can be derived in the asymptotic limit, where $D\xrightarrow{}\infty$~\cite{Ostrowski}.

Within the $\lambda$-family, a minimum value of $Q=-0.5$ is attained when $\lambda=0.5$, which defines the matrix elements of the gain and loss operators as reciprocals to one another. This value for $Q$ corresponds to a $50\%$ reduction below the shot noise limit. With the $q$-family, we instead find that $100\%$ noise reduction in the beam is achievable by imposing a completely regular pump with $q\xrightarrow{}-1$. Creating a sub-Poissonian beam in this manner, by means of subjecting the laser to a pump which itself is sub-Poissonian, is well known~\cite{Walls_Milb,Drummond}; it is also known that imposing such a pump in an otherwise standard laser has no effect on the rate of phase diffusion (and hence coherence)~\cite{Bergou,Benkert_Phase,wiseman1993,Drummond}. What is interesting about the results at hand is that the models we study exhibit a phase diffusion vastly smaller than standard, and for both families, sub-Poissonian statistics in the output field ensue when measures are taken to increase the coherence. In other words, we find that there is a ``win-win relationship" between coherence and sub-Poissonianity for optimized Heisenberg-limited lasers.

In this Letter we have shown how the Heisenberg limit for laser coherence may be generalized to encompass beams that may be highly sub-Poissonian. From this result, we have found that reducing the photon noise for such a beam can in fact be advantageous for a reduction in the phase noise. This marks a generalization of past studies into sub-Poissonian lasers with standard (linear) photon loss, to beams which exhibit phase noise that is as small as permitted by the laws of quantum mechanics. This work could have applications in technologies requiring minimized noise in both intensity and phase, such as quantum information processing.

We acknowledge the support of the Griffith University eResearch Service and Specialised Platforms Team, and the use of the High Performance Computing Cluster ``Gowonda" to complete this research. This research was funded by the Australian Government through the through the Australian Research Council's Discovery Projects funding scheme (project DP220101602), and an Australian Government Research Training Program (RTP) Scholarship.


\end{document}